\documentclass[showpacs,prd,twocolumn,amsmath,amssymb,longbibliography]{revtex4-1}
\usepackage{graphicx}
\usepackage{amsmath,amsfonts}
\usepackage{dcolumn}
\usepackage{bm}
\usepackage{slashed}
\usepackage{nicefrac}
\usepackage{epstopdf}
\usepackage{color}
\usepackage{tikz-feynman,contour}
\def\bra{\langle}
\def\ket{\rangle}
\def\cD{\mathcal{D}}
\def\cN{\mathcal{N}}
\def\cH{\mathcal{H}}
\def\lr{{L^2(\R^d)}}
\def\vq{\mathbf{q}}
\def\vy{\mathbf{y}}
\def\R{\mathbb{R}}
\def\Tr{\mathrm{Tr}}
\def\Ei{\mathrm{Ei}}
\def\dk#1#2{\frac{ d^{#2}{#1} }{ (2\pi)^{#2} }} 

\begin{document}
\title{Are there any Landau poles in wavelet-based quantum field theory?}
\author{Mikhail Altaisky} 
\affiliation{Space Research Institute RAS, Profsoyuznaya 84/32, Moscow, 117997, Russia}
\affiliation{School of Energy Materials, Mahama Gandhi University,Priyadarsini Hills, Kottayam, Kerala, 686560, India}
\email{altaisky@rssi.ru}

\author{Michal Hnatich} 
\affiliation{Bogoliuobov Laboratory of Theoretical Physics, Joint Institute for Nuclear Research, \\ Joliot-Curie 6, 
Dubna, 141980, Russia}
\affiliation{P.J.\v{S}af\`{a}rik University in Ko\v{s}ice, Park Angelinum 9, 04154, Ko\v{s}ice, Slovakia}
\affiliation{Institute of Experimental Physics SAS, Watsonova 47, 04001, Ko\v{s}ice, Slovakia}
\email{hnatic@saske.sk}
\date{October 4, 2023}

\begin{abstract}
Following previous work by one of the authors [M.V.Altaisky, Unifying renormalization group and the continuous wavelet transform, Phys. Rev. D 
{\bf 93}, 105043 (2016).], we develop a new approach to the renormalization group, where the effective 
action functional $\Gamma_A[\phi]$ is a sum of all fluctuations of scales from the size of the system $L$
down to the scale of observation $A$. It is shown that the 
renormalization flow  equation of the type    
$
\frac{\partial \Gamma_A}{\partial \ln A}=-Y(A)
$ is a limiting case of such consideration, when the running coupling constant is assumed 
to be a differentiable function of scale. In this approximation, the running coupling constant, 
calculated at one-loop level, suffers from the Landau pole. In general case, when the 
scale-dependent coupling constant is a non-differentiable function of scale, the Feynman loop expansion results in a difference equation. This keeps 
the coupling constant finite for any finite value of scale $A$. As an example we 
consider Euclidean $\phi^4$ field theory. 
\end{abstract}
\maketitle
\section{Introduction \label{intro:sec}} 
The renormalization group (RG) was discovered by Stueckelberg and Petermann as a group of parametrizations
of the $S$-matrix emerging after cancellation of UV divergences in quantum electrodynamics \cite{SP1953}. The RG method has become popular 
in high-energy physics since Gell-Mann and Low, who used the functional equation to study the renormalized photon propagator in QED, have shown that the charge distribution surrounding a test charge in vacuum does not depend on the coupling 
constant at small scales, except for a scale factor, i.e., it posses a kind of self-similarity \cite{GL1954}.
The breakthrough in  the RG approach has been achieved by Wilson, who applied it to statistical mechanics, 
where continuously many degrees of freedom are correlated over long distances. It was found that if there are 
many degrees of freedom  within the correlation length $\xi$, the behavior of the system is primarily determined by 
cooperative behavior and the number of degrees of freedom, rather than by the type of 
Hamiltonian interaction \cite{WK1974}. The core of the Wilson formulation was to successively integrate out 
the fluctuations of small scales to obtain progressively coarse-grained descriptions of fluctuations at 
larger scales \cite{Wilson1971a,Wilson1971b}.
By doing so, the RG approach unifies the theory of phase transitions, quantum field theory, turbulence and 
many other branches of physics. The RG approach not only provides an explanation of critical phenomena, but also 
renders a practical tool for calculation of second-order phase transitions \cite{WK1974,PELISSETTO2002}.

Having started from the weak interaction limit, where one-loop corrections to the correlation functions have 
tractable physical meaning, the RG approach has gradually evolved into the scaling equations for the exact (''dressed'') correlation functions. In this paper, following the previous papers \cite{Altaisky2020PRD,Altaisky2016PRD,AR2020}, we sum up the fluctuations starting from the IR edge and go 
down to the observation scale. If the fluctuations are summed up in a thin shell of scales, the beta function coincides 
with the known results, regardless the summation starts from the IR or from the UV edge \cite{Altaisky2016PRD}. 
We have shown that summing up the fluctuations from IR edge (from size of the system) down 
to the observation scale in a finite range of scales, renders a finite renormalization of the coupling 
constant without any Landau poles.

The use of {\em continuous} wavelet transform is not the only way of the wavelet regularization in quantum field 
theory aimed to sum up the fluctuations of different scales. 
Historically, the first attempts to use wavelets in quantum field theory were related to the numerical 
simulation of QFT models. For instance, Ref. \cite{HS1995} presents a simple 2D model with local 
$\phi^4$ interaction simulated using discrete wavelet transform (DWT) of the field $\phi(x)$ performed with 
orthogonal Daubechies wavelets. The Daubechies wavelets  \cite{Daub1988} form an orthogonal 
basis of the compactly supported functions $\psi^j_k(x):=2^{-\frac{j}{2}}\psi(2^{-j}x -k)$, where $\psi(x)$ is a 
function with compact support, obeying certain recursive equation.
The advantage found in wavelet simulations was that the coefficients of different scales $j$ were varied {\em independently} when searching for ground state of the field configuration. This effectuates 
the simulation if compared to the usual Metropolis algorithm.

The main advantages of DWT in quantum field theory simulations are more or less the same as its advantages in signal 
and image processing, where independence of coefficients of different scales provides fast and efficient algorithms 
for data processing \cite{Daub10,Beylkin1992}. As it concerns the  quantum field theory itself, the technique based 
on DWT was later generalized to the {\em multiscale entanglement renormalization ansatz}, closely related to 
real space renormalization group \cite{Vidal2007,Vidal2008}. Wavelet technique is also closely related to the RG, since 
the wavelet basis is generated by a single basic function, which is shifted and dilated to form the bases of different scales. This scale hierarchy of wavelet bases provides a natural framework for renormalization on a lattice  \cite{EW2016,Haegeman2018,SMMT2021}. 

There is essential difference between usual renormalization procedures -- Kadanoff blocking procedure \cite{Kadanoff1966}, Wilson's RG \cite{Wilson1971a},  
etc, -- and the wavelet technique. In the usual approach there is only one universal operator $D$, which makes the 
blocks of different sizes behave like each other, although with different values of the coupling constant. 
In discrete wavelet transform there are two distinct operators: the low-pass and high-pass filters. The 
former averages degrees of freedom of the small scales into the coarser degrees of freedom of the block -- exactly 
as in usual RG approach -- the latter process the details lost in block averaging, $\hat{l}+\hat{h}=\hat{\mathbb{I}}$,  \cite{Daub10,EW2016}. For the case of discrete wavelet transform, the problem of lattice renormalization was described in detail in the monograph \cite{Battle1999}.  

In this paper, we are not going to dig into details of DWT methods in quantum field theory, but we do  
mention that these methods have gradually evolved into an effective numerical technique for finding the ground 
state of many-body quantum field theory models \cite{Brennen2022}. Their promising implementation 
is arising due to the analogy between DWT and the tensor networks 
\cite{CV2009,Singh2018} is expected on quantum computers \cite{Brennen2015}. Discrete wavelet representation of 
field theory models with the orthogonal Daubechies wavelets also provides an interesting approach to the 
evolution of lattice Hamiltonian systems \cite{BP2013,Polyzou2020,Brennen2022}.

To the authors knowledge, the first attempt to use wavelets for an analytic study of a continuous field theory 
of practical importance -- the quantum chromodynamics beyond the lattice approximation -- was made by 
 Federbush \cite{Federbush1995}, but has turned out to be technically complicated. In this study the basis 
was not restricted to the orthogonal wavelets, and the general type of DWT was considered, with the examples of Meyer wavelets. The mathematical idea itself was expressed even earlier \cite{BF1983}.

In our approach, following the previous papers \cite{Altaisky2010PRD,AK2013,Altaisky2016PRD} we use 
{\em continuous} wavelet transform, rather than {\em discrete} wavelet transform. By doing so, we lose orthogonality of the basis, but harness the capability of analytical calculations in the perturbation theory  
and shape our model into the framework of the group field theory \cite{ffp4,Freidel2005}, defined on the 
affine group $G:x' = ax+b$. The use of continuous wavelet transform provides a quantum field theory model 
finite by construction, if the scale argument of wavelet transform ($a$) is considered as a physical scale of measurement 
of a quantum field performed with a certain aperture function -- mother wavelet -- and the causality 
restriction being imposed on scale arguments, as the absence of scales in the internal lines of Feynman diagrams 
less than the best scale of measurement, given as minimal scale of all external lines \cite{Alt2002G24,Altaisky2010PRD}. In this settings the scale dependence of the observed Green functions (field 
correlators $\langle \phi_{a_1}(x_1)\ldots \phi_{a_n}(x_n)\rangle$) is completely expressed in terms of the wavelet scale arguments $a_i$, no external renormalization 
scale $\mu$ is required, and the role of renormalization group symmetry is merely to relate the fluctuations of 
different scales to each other: there is no need in removal of divergences \cite{Altaisky2016PRD}.

The remainder of this paper is organised as follows. In {\em Sec. \ref{frg:sec}} we briefly remind the reader of 
some definitions of the Euclidean field theory in its statistical interpretation. {\em Section \ref{cwt:sec}} 
presents the formalism of continuous wavelet transform in Euclidean QFT. In {\em Sec. \ref{phi4:seq}} we 
present one-loop contribution to the vertex in $\phi^4$ model, and show that accurate summation of contributions 
of all scales from the size of the system down to the observation scale does not produce any singularities such 
as Landau poles. A few concluding remarks are given in the last section.

\section{Statistical mechanics view on quantum field theory\label{frg:sec}} 
Let us briefly remind the statistical view on the formalism of Euclidean quantum field theory. 
At the state of thermodynamic equilibrium the distribution of a continuous field $\phi(x)$, say a magnetization, is  
given by the canonical partition function
$$
Z = \Tr e^{-\beta H},
$$
where $H=H[\phi]$ is the Hamiltonian, $\beta\!=\!\frac{1}{T}$ is the inverse temperature, and the trace 
operator assumes the summation over all degrees of freedom. The trace can be expressed in terms of the 
Feynman integral,
\begin{equation}
Z[J]= \int \cD \phi \exp\left(-S[\phi]+\int J(x)\phi(x)d^dx \right), \label{qft1}
\end{equation}
where the formal source $J(x)$ can be understood as an external magnetic field.

The Euclidean action functional $S[\phi]$ is proportional to the Hamiltonian of the field $\phi(x)$, 
\begin{equation}
S[\phi] = \frac{1}{T}\int d^dx \left[ \frac{1}{2} (\partial\phi)^2 + \frac{m^2}{2} \phi^2  
+ \frac{\lambda}{4!}\phi^4\right], \label{se4}
\end{equation}
in the Ginzburg-Landau theory of phase transitions \cite{GL1950}.
The correlation functions of the field $\phi(x)$ can be derived as functional derivatives, 
\begin{equation}
G^{(n)}(x_1,\ldots,x_n) = 
\left. { \frac{\delta^n W[J]}{\delta J(x_1) \ldots \delta J(x_n)}
}\right|_{J=0} \label{cgf}, 
\end{equation} 
where $W[J] = \ln Z[J]$ is the connected Green functions generating functional, which is proportional 
to the Helmholtz free energy $F[J]=-T \ln Z[J]$.
The effective action functional $\Gamma[\phi]$ is defined via the Legendre transform of 
$W[J]$,
\begin{equation}
\Gamma[\phi] =   -W[J] + \int J(x) \phi(x) d^dx . \label{eaf}
\end{equation}
(Here we keep the notation of \cite{BTW2002}.) 

The functional derivatives of $W[J]$ with respect to the external source $J(x)$ determine the  mean field 
$\phi = \phi[J]$:
$$
\frac{\delta W[J]}{\delta J(x)} = \phi(x).
$$
The functional derivatives of the effective action $\Gamma[\phi]$ are the {\em vertex functions} $\Gamma^{(n)}[\phi]$. 
In the above considered $\phi^4$ model, the (renormalized) vertex function  $\Gamma^{(4)}[\phi]$ accounts for the 
value of coupling constant calculated at some reference scale; the $\Gamma^{(2)}[\phi]$ function is the 
renormalized inverse propagator, which defines the renormalization of mass at the same reference scale.

The most instructive case of the locally known microscopic interaction is the Ising model, 
described by microscopic Hamiltonian 
\begin{equation}
H = -J \sum_{<ij>}S_i S_j - B \sum_i S_i, \label{ising}
\end{equation}
where $J$ is the coupling constant of interaction between the neighbouring spins, $B$ is external 
magnetic field, and the Ising spins, with the values $S_i=\pm 1$, are located on some regular lattice. In continuous 
limit, the Hamiltonian of the Ising model \eqref{ising} with the nearest-neighbor interaction is transferred into Euclidean QFT model with 
$\phi^4$ interaction \eqref{se4}, which meets the Ginzburg-Landau theory \cite{GL1950,WK1974}.

In many cases, the interaction Hamiltonian or the bare action functional is known at some 
{\em macroscopic} scale $\mu$, but the microscopic theory at smaller scales (higher momentum transfer) 
should be unveiled. The typical cases are the QED and the quantum gravity -- both having $1/r$ asymptotic behaviour at macroscopic scales, but different behaviour at smaller scales \cite{Reuter1998,Reuter2009}.

The renormalization group method displays its best merits when the microscopic fluctuations of atomic scales 
cooperate their behavior into large-scale fluctuations, which are well described by classical mean-field 
equations. This happens in the theory of phase transitions, critical behavior, kinetic description of gases, 
etc. \cite{WK1974,Vasiliev1998,McComb2004}.
However, if fluctuations of all scales do matter equally, the averaging of fluctuations from the atomic scales up to 
the larger scales (say, by  Bogolubov's chain) becomes notoriously difficult. It turns easier, say in hydrodynamics, to start 
with the laminar large  scale motion and to sum up all fluctuations arising from instabilities down to the 
atomic scales, where these fluctuations are completely damped by viscosity \cite{Vasiliev1998}.

\section{Using continuous wavelet transform in quantum field theory models \label{cwt:sec}}
\subsection{Continuous wavelet transform}
To separate fluctuations of different scales in quantum field theory,  it is convenient to use the formalism of continuous wavelet transform (CWT), 
as described,  e.g., in \cite{Alt2002G24,Altaisky2010PRD}. Let us briefly remind the basics of wavelet transform, 
see the monographs \cite{Daub10,Chui1992} for a detailed introduction. 

Let $\phi(x)\in \lr$ be a square-integrable function. Let $\chi(x)\in\lr$ be a suitably well-localized function,
which satisfies the {\em admissibility condition}
\begin{equation}
C_\chi = \int |\tilde{\chi}(k)|^2 \frac{d^dk}{S_d |k|^d}<\infty, \label{adcf}
\end{equation}
where tilde denotes the Fourier transform,
$$
\tilde{\chi}(k):=\int_{\R^d} e^{\imath k x} \chi(x) d^dx,$$
and 
$\quad S_d = \frac{2 \pi^{d/2}}{\Gamma(d/2)}$ is the area of unit sphere 
in $\R^d$, 
then it is possible to decompose the function $\phi$ with respect to the basis, provided by shifted, 
dilated, and rotated copies of $\chi(x)$. This decomposition is known as continuous wavelet transform (CWT) 
\cite{GGM1984,GM1984}:
\begin{equation}
\phi(x) = \frac{1}{C_\chi} \int \frac{1}{a^d} \chi\left(R^{-1}(\theta)\frac{x-b}{a}\right) \phi_{a\theta}(b) \frac{dad^db}{a} d\mu(\theta), \label{iwt}
\end{equation} 
where $R(\theta)$ is the rotation matrix, $d\mu(\theta)$ is the left-invariant measure on the $SO(d)$ rotation group, usually written in terms of the Euler angles, 
$$
d\mu(\theta) = 2\pi \prod_{k=1}^{d-2} \int_0^\pi \sin^k \theta_k d\theta_k.
$$
The functions 
\begin{equation}
\phi_{a,\theta}(b) := \int_{\R^d} \frac{1}{a^d} \overline{\chi} \left(R^{-1}(\theta)\frac{x-b}{a} \right) \phi(x) d^dx \label{dwtrd}
\end{equation} 
are known as wavelet coefficients of the function $\phi$ with respect to the mother wavelet $\chi$.

The decomposition \eqref{dwtrd} and the reconstruction formula \eqref{iwt}  represent a particular case of the 
''partition of unity'' in Hilbert space $\cH$ with respect to representation $U(g)$ of a Lie group $G$ acting transitively on $\cH$ \cite{Carey1976,DM1976},
\begin{equation}
\hat{\mathbb{I}} = \frac{1}{C_\chi} \int_G U(g)|\chi\rangle d\mu(g) \langle \chi |U^*(g),
\label{pu:eq}
\end{equation}
with $G$ being the group of affine transformations,
\begin{equation}
G: x' = a R(\theta)x + b, x,b \in \R^d, a \in \R_+, \theta \in SO(d). \label{ag1}
\end{equation}

Wavelet coefficients \eqref{dwtrd} have clear physical meaning: The convolution of the analyzed function 
$\phi$ with  a well-localized function $\chi$ at a fixed window width $a$ comprise only the fluctuations with 
typical scales close to $a$  and is insensitive to all other fluctuations. 

\subsection{Scale-dependent fields}
The reconstruction \eqref{iwt} of the function $\phi$ from the set of its wavelet coefficients is generally 
non-orthogonal, and the wavelet basis is overcomplete \cite{Chui1992}. Although the integration 
$\int_0^\infty \frac{da}{a} \ldots $ in \eqref{iwt} provides a formally exact reconstruction formula,
depending on the physics of the considered problem, we can restrict the integration by the minimal scale 
$A$ from below (lattice size -- in the case of ferromagnetism) and by the system size $L$ from above: 
$\int_A^L \frac{da}{a} \ldots$

Moreover, as we know from the Heisenberg uncertainty principle, the value of a {\em quantum} field $\phi$ 
sharp at a point $x$ is physically meaningless, since any measurement with $\Delta x\!\to\!0$ implies an 
infinite momentum transfer $\Delta p\!\to\infty$, which definitely drives us out of the applicability of the model.
For this reason we have to consider $A$ as the best available scale of measurement (observation).

There is a distinction between quantum field theory models considered as an effective large scale 
description of a microscopic theory with a fundamental microscopic length scale -- say, a ferromagnetic model -- 
and fundamental models of quantum field theory. Our approach is oriented for the latter case. 
In the former case we have a physical evidence of what is happening at the fundamental scale, and the goal 
of RG or any other technique is to construct an effective large scale theory by a certain averaging procedure.
In the latter case our physical evidence is related to some large scale processes -- say, the interaction of charged 
particles in classical electrodynamics  -- and our goal is to construct a microscopic theory capable of 
describing physical phenomena at a given microscopic scale of observation. We know from theoretical calculations 
in QED, which harness RG technique in the space of local square-integrated functions $\mathrm{L}^2(\R^d)$, 
that some experimental results, such as Lamb shift and anomaly in magnetic momentum can be explained in this 
way \cite{Schwinger1946,Dehmelt1987}.

Nevertheless, we do not have any proof that the description of quantum fields in terms of local square-integrable fields and constant charges is the unique way to describe quantum phenomena. The techniques of continuous wavelet 
transform, as presented in \cite{Altaisky2010PRD} and other papers, suggests an alternative description: quantum 
fields may be defined on affine group $G: x'=ax+b$, rather than on Euclidean or Minkowski space: $\phi=\phi_a(b)$.
In this case the charges may be explicitly dependent on scale $Q=Q(a)$.  In such a theory the no-scale (''bare'') coupling constant may have no 
physical meaning, but we still keep it to link with the known results. There is a counterpart for this situation  
in classical statistical mechanics and turbulence theory. The definition of viscosity 
and other kinetic coefficients is explicitly dependent on the size of averaging domain; and there is nothing strange 
in it: if the averaging size is less than the mean free path, neither viscosity, nor hydrodynamic velocity is defined, which drives us out of the model applicability. The continuous wavelet transform can be also applied 
to such problems \cite{AHK2018}.


In the remainder of this paper, following the previous papers \cite{Altaisky2010PRD,Altaisky2020PRD} we will assume the mother wavelet $\chi(x)$ to be an isotropic function 
of $x$ and thus ignore the rotation factor $R(\theta)$. In this settings, the scale component of the field $\phi$, measured 
in a point $b$ at the scale  $a$ with respect to the mother wavelet $\chi$ (considered as an aperture function by the analogy from optics \cite{PhysRevLett.64.745}) is given by wavelet 
coefficient,
\begin{equation}
\phi_a(b)\equiv \bra a,b;\chi|\phi\ket = \int \frac{1}{a^d} \overline{\chi} \left(\frac{x-b}{a} \right) \phi(x) d^dx
.
\label{cwti}
\end{equation}
However, the space of scale-dependent functions $\{ \phi_a(b) \}$ is more general than the space of point-dependent 
functions $\phi(x)\in\lr$. Even if all fields $\phi_a(b)$ are well defined $\forall a\in \R_+, b\in \R^d$, the 
limit 
$$
\phi(x) = \lim_{A\to0} \int_A^\infty \frac{da}{a}\int_{\R^d} \frac{1}{a^d} \chi \left(\frac{x-b}{a} \right) \phi_a(b) d^dx
$$
does not necessarily exist. The divergence of the sum of all scale components happens in UV-divergent theories, 
where the value of a physical field $\phi$ sharp at a point $x$ is meaningless.

If $\phi(x)$ is understood as a wave function of physical particle, its normalization 
$$
\bra \phi | \phi \ket = \int \bar{\phi}(x) \phi(x)d^dx =1
$$ 
is the statement of existence: the probability of finding this particle anywhere in space $\R^d$ is exactly 1.
The wavelet approach, based on the affine group \eqref{ag1}, generalizes the statement of existence in the form 
\begin{equation}
\frac{1}{C_\chi} \int_{g\in G} |\bra\phi|U(g)|\chi \ket|^2 d\mu(g) = 1, \quad g=(a,b,\theta).
\end{equation}
The latter equation states that sweeping the measurement parameters, position $b$ and the resolution $a$ and the direction $\theta$, over all 
possible values will necessarily imply the registration of the particle.

Technically, the use of the scale-dependent functions $\phi_a(b)$ in a local quantum field theory is rather straightforward: one can express local fields in terms of their wavelet transform, 
\begin{equation}
\phi(x) = \frac{1}{C_\chi} \int \frac{da}{a} \int \frac{d^dk}{(2\pi)^d}e^{-\imath k x} \tilde{\chi}(ak)
\tilde{\phi}_a(k),
\end{equation}
where $\tilde{\phi}_a(k)$ are the Fourier images of the wavelet coefficients \eqref{cwti}.
This defines an easy rule to redefine the Feynman diagram technique,
\begin{equation}
\tilde{\phi}(k) \to \tilde{\phi}_a(k) = \overline{\tilde{\chi}(ak)}\tilde{\phi}(k). \label{r1}
\end{equation} 
Doing so, we have the following modification of the Feynman diagram technique
\cite{Alt2002G24,Altaisky2020PRD}:
\begin{enumerate}\itemsep=0pt
\item Each field $\tilde\phi(k)$ is substituted by the scale component: 
$\tilde\phi(k)\to\tilde\phi_a(k) = \overline{\tilde \chi(ak)}\tilde\phi(k)$.
\item Each integration in the momentum variable is accompanied by the corresponding 
scale integration,
\[
 \dk{k}{d} \to  \dk{k}{d} \frac{da}{a} \frac{1}{C_\chi}.
 \]
\item Each interaction vertex is substituted by its wavelet transform; 
for the $N$-th power interaction vertex, this gives multiplication 
by the factor 
$\displaystyle{\prod_{i=1}^N \tilde \chi(a_ik_i)}$.
\end{enumerate}
According to these rules, the bare Green function of a massive scalar field in wavelet representation 
takes the form
$$
G^{(2)}_0(a_1,a_2,p) = \frac{\tilde \chi(a_1p)\tilde \chi(-a_2p)}{p^2+m^2}.
$$    
The finiteness of loop integrals is provided by the following rule:
{\em There should be no scales $a_i$ in internal lines smaller than the minimal scale 
of all external lines} \cite{Alt2002G24,Altaisky2010PRD}. Therefore, the integration in $a_i$ variables is performed from the minimal scale of all external lines up to infinity or up to the system size. 
This corresponds to the summation of all 
fluctuations of all scales from the system size down to the fines scale of observation.

The cutoff in scale variables $a$ is a milder assumption than momentum cutoff $\Lambda$ in a usual theory.
Since the scale $a$ is a setting of observation, rather than a measurable quantity like momentum, the 
cutoff in it results neither in violation of momentum conservation, nor in violation of other important symmetries. 

For a theory with local $\phi^N(x)$ interaction, the presence of two conjugated factors $\tilde{\chi}(ak)$ 
and $\overline{\tilde{\chi}(ak)}$ on each diagram line connected to interaction vertex, simply means 
that each internal line of the Feynman diagram carrying momentum $p$ is supplied by the cutoff factor 
$f^2(Ap)$, where 
\begin{equation}
f(x) := \frac{1}{C_\chi} \int_x^\infty |\tilde{\chi}(a)|^2 \frac{da}{a}, \quad f(0)=1, \label{fcut}
\end{equation}
with $A$ being the minimal scale of all external lines of this diagram.

The mildness of the cutoff in scale argument $A$, if compared to the momentum cutoff $\Lambda \sim A^{-1}$, is quite easy to understand. 
Let us take the $\phi^4$ model and consider a ''fish'' diagram 
\begin{equation}
\begin{tikzpicture}[baseline=(e)]
  \begin{feynman}[horizontal = (e) to (f)]
  \vertex [dot] (e);
  \vertex [dot,right=1.5cm of e] (f);
  \vertex [above left = 1cm of e] (i1) {\(p_1\)};
  \vertex [below left = 1cm of e] (i2) {\(p_2\)};
  \vertex [above right = 1cm of f] (i3) {\(p_3\)};
  \vertex [below right = 1cm of f] (i4) {\(p_4\)};
  \diagram*{
  {(i1),(i2)} --  (e),
  {(i3),(i4)} --  (f),
  (e) -- [half right, edge label = \(q_2\)] (f),
  (e) -- [half left, edge label = \(q_1\)] (f),
    };
  \end{feynman} 
  \end{tikzpicture}
\label{fish}
\end{equation}
with the loop momenta $q_1$ and $q_2$ satisfying the momentum conservation in both vertices. 
In case of the Fourier decomposition of fields, the restriction of momentum integration by 
cutoff momentum $\Lambda$ results in low-frequency fields, 
$$
\phi^{<}_\Lambda(x) := \int_{|q|<\Lambda} e^{-\imath q x} \tilde{\phi}(q) \dk{q}{d}.
$$
If both $q_1$ and $q_2$ are above the cutoff value $\Lambda$, the contribution of both components 
to the loop integral will be discarded. More than that, since the Fourier transform is a decomposition 
with respect to the representations of translation group, the application of cutoff violates translational 
invariance. Since the momentum basis is {\em orthogonal} basis, 
$$\hat{\mathbb{I}} = \int |k\ket \dk{k}{d} \bra k|, \quad \bra k|k'\ket=(2\pi)^d\delta(k-k'),$$ 
some information is lost after cutting high momenta. 

In contrast to that, the wavelet basis is generally non-orthogonal and overcomplete, and in discrete case forms a frame, see, e.g., \cite{Chui1992} for 
general introduction to wavelets. In continuous case wavelet transform is a decomposition of a function 
with respect to representations of affine group 
$
G: x'=ax+b,
$
see Eq.\eqref{pu:eq}. 
When we restrict the integration over 
a scale domain $A \le a < \infty$, the translation subgroup ($b$) is not affected, and the translation invariance is preserved. For the ''fish'' diagram \eqref{fish}, 
after application of scale cutoff $A$ , both components, with momenta $q_1$ and $q_2$ will contribute, but their 
contributions will be modulated by $f^2(Aq_1)$ and $f^2(Aq_2)$, respectively. The analogues of these contributions 
are completely lost in usual Fourier method.

This is a typical story in information theory, when introduction of a new dimension -- scale $a$ in our case --
enables one to preserve more information than available for usual methods. In machine learning, this stimulates the 
use of higher-dimensional feature maps \cite{SP2018}. For the same reasons, wavelets benefit in signal processing, 
where they are capable of distinguishing the low frequencies coming from the differences of two high-frequency 
harmonics from those coming from a natural low-frequency source \cite{GGM1984}.

The summation of all fluctuations from the system size down to the finest observation scale, but not below it, 
seems quite natural, for the integration over infinitely small scales is often beyond the applicability 
range of a particular physical model. This happens in ferromagnetic theories below the grid spacing, in turbulence below 
the mean free path, etc.

\subsection{Mother wavelets}
\begin{figure}[t]
\centering \includegraphics[width=8cm]{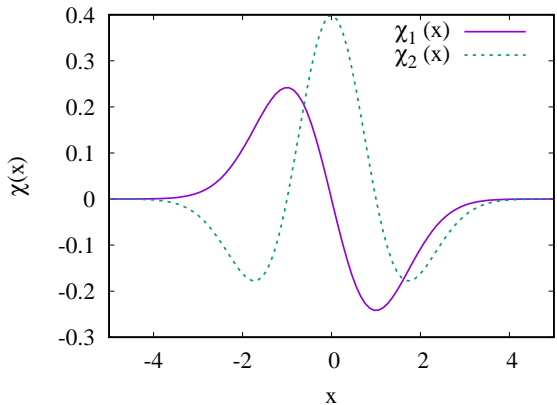}
\caption{First two wavelets of the wavelet family \eqref{gn}}
\label{g12:pic}
\end{figure}

In our calculations, we use different derivatives of the Gaussian as mother wavelets. The admissibility condition \eqref{adcf} is rather loose: practically any well-localized function with the Fourier image vanishing at zero momentum ($\tilde{\chi}(0)\!=\!0$) obey this requirement. As for the Gaussian functions, 
\begin{equation}
\chi_n(x) = (-1)^{n+1} \frac{d^n}{dx^n} \frac{e^-\frac{x^2}{2}}{\sqrt{2\pi}},\quad n>0,
\label{gn}
\end{equation}
where $x$ is a dimensionless argument,
they are easy to integrate in Feynman diagrams. The graphs of first two wavelets of the \eqref{gn} family, 
$$
\chi_1(x) = -\frac{x e^{-\frac{x^2}{2}}}{\sqrt{2\pi}}, \quad \chi_2(x) = \frac{(1-x^2) e^{-\frac{x^2}{2}}}{\sqrt{2\pi}},
$$
are shown in Fig.~\ref{g12:pic}. 
Their Fourier images are 
\begin{equation}
\tilde{\chi}_n(k) = - (\imath k)^n e^{-\frac{k^2}{2}}.  
\end{equation}
Respectively, the normalization constants and the wavelet cutoff functions are
\begin{align*}
C_{\chi_n} = \frac{\Gamma(n)}{2}, \quad f_{\chi_n}(x) = \frac{\Gamma(n,x^2)}{\Gamma(n)},
\end{align*}
where $\Gamma(\cdot)$ is the Euler gamma function, and $\Gamma(\cdot,\cdot)$ is the incomplete gamma function. 
For the first two wavelets of the family \eqref{gn} the  cutoff functions are:  
\begin{equation}
f_{\chi_1}(x) = e^{-x^2}, \label{fch1}\quad  f_{\chi_2}(x) = (1+x^2) e^{-x^2}. 
\end{equation}

\section{An example of $\phi^4$ model \label{phi4:seq}}
Let us consider an Euclidean action functional of a massive scalar field with the local $\phi^4$ interaction \eqref{se4},
$$
S[\phi] = \int d^dx \left[ \frac{1}{2} (\partial\phi)^2 + \frac{m^2}{2} \phi^2  
+ \frac{\lambda}{4!}\phi^4\right].
$$
This model is an extrapolation of a classical interacting spin model to the continual limit \cite{GJ1981}. Known as the Ginzburg-Landau 
model \cite{GL1950}, it describes phase transitions in superconductors and other  magnetic systems fairly well, but it produces divergences when the 
correlation functions 
are evaluated from the generating functional \eqref{qft1}
by perturbation expansion; see, e.g., \cite{ZJ1999} for a discussion.

The parameter $\lambda$ in the action functional \eqref{se4} is a phenomenological coupling constant, which knows nothing about 
the scale of observation and becomes the running coupling constant only because of renormalization or the cutoff 
application. The straightforward way to introduce scale dependence into the local model \eqref{se4} is to express 
local field $\phi(x)$ in terms of its scale components $\phi_a(b)$ using the inverse wavelet transform \eqref{iwt}.
This leads to the generating functional for scale-dependent fields,
\begin{widetext} 
\begin{align} \nonumber 
Z_W[J_a] &=&\cN \int \cD\phi_a(x) \exp \Bigl[ -\frac{1}{2}\int \phi_{a_1}(x_1) D(a_1,a_2,x_1-x_2) \phi_{a_2}(x_2)
\frac{da_1d^dx_1}{C_\chi a_1}\frac{da_2d^dx_2}{C_\chi a_2}  \\
&-&\frac{\lambda}{4!}
\int V_{x_1,\ldots,x_4}^{a_1,\ldots,a_4} \phi_{a_1}(x_1)\cdots\phi_{a_4}(x_4)
\frac{da_1 d^dx_1}{C_\chi a_1} \frac{da_2 d^dx_2}{C_\chi a_2} \frac{da_3 d^dx_3}{C_\chi a_3} \frac{da_4 d^dx_4}{C_\chi a_4} 
+ \int J_a(x)\phi_a(x)\frac{dad^dx}{C_\chi a}\Bigr], \label{gfw}
\end{align}
\end{widetext}
where $D(a_1,a_2,x_1-x_2)$ is the wavelet transform of the ordinary propagator, and $\cN$ is a formal normalization constant \cite{Alt2002G24}.

The functional \eqref{gfw} -- if integrated over all scale arguments in infinite limits $\int_0^\infty \frac{da_i}{a_i}$ -- will certainly drive us back to the known divergent theory \eqref{qft1}. All scale-dependent fields $\phi_a(x)$ in 
Eq.\eqref{gfw} still interact with each other with the same coupling constant $\lambda$, but their interaction is now 
modulated  by the wavelet factor $V_{x_1x_2x_3x_4}^{a_1a_2a_3a_4}$, which is the Fourier transform of 
$\prod_{i=1}^4 \tilde{\chi}(a_ik_i)$.

As usual in functional renormalization group technique \cite{Wetterich1993}, we can introduce the effective action 
functional \eqref{eaf},
the functional derivatives of which are the vertex functions $\Gamma_{(A)}^{(n)}$,
\begin{widetext}
$$
\Gamma_{(A)}[\phi_a] = \Gamma_{(A)}^{(0)} + \sum_{n=1}^\infty \int  
\Gamma_{(A)}^{(n)}(a_1,b_1,\ldots,a_n,b_n) \phi_{a_1}(b_1)
\ldots \phi_{a_n}(b_n) \frac{da_1d^db_1}{C_\chi a_1}
\ldots \frac{da_nd^db_n}{C_\chi a_n}.
$$
\end{widetext}
The subscript $(A)$ indicates the presence in the theory of 
some minimal scale -- the observation scale. 

In one-loop approximation, the two-point and the four-point vertex functions, $\Gamma^{(2)}$ and 
$\Gamma^{(4)}$, respectively, are given by the following diagrams:
\begin{equation}
\Gamma^{(2)} = \Delta_{12} - \frac{1}{2}
\begin{tikzpicture}[baseline=(a)]
\begin{feynman}[inline=(a)]
\vertex [blob] (a);
\node [left=1cm of a]  (i1){\(1\)};
\node [right=1cm of a] (i2){\(2\)};
\vertex [above=1cm of a] (b);
\vertex [above=1.5cm of a] (c);
\diagram[horizontal=i1 to i2]{
(i1) -- (i2),
(a) -- [half left] (b),
(b) -- [half left] (a),
};
\end{feynman}
\end{tikzpicture} \label{G2:fde}
\end{equation}
\begin{equation}
\Gamma^{(4)} = -
\begin{tikzpicture}[baseline=(a)]
\begin{feynman}[inline=(a)]
    \vertex [dot] (a);
    \vertex [above=1cm of a] (i1){\(1\)};
    \vertex [left =1cm of a] (i2){\(2\)};
    \vertex [right=1cm of a] (i3){\(3\)};
    \vertex [below=1cm of a] (i4){\(4\)};
    \diagram*{  
    {(i1),(i2),(i3),(i4)} --  (a),
     };
\end{feynman}      
\end{tikzpicture}
  -\frac{3}{2}
  \begin{tikzpicture}[baseline=(e)]
  \begin{feynman}[horizontal = (e) to (f)]
  \vertex [dot] (e);
  \vertex [dot,right=1.5cm of e] (f);
  \vertex [above left = 1cm of e] (i1) {\(1\)};
  \vertex [below left = 1cm of e] (i2) {\(2\)};
  \vertex [above right = 1cm of f] (i3) {\(3\)};
  \vertex [below right = 1cm of f] (i4) {\(4\)};
  \diagram*{
  {(i1),(i2)} --  (e),
  {(i3),(i4)} --  (f),
  (e) -- [half right] (f),
  (e) -- [half left] (f),
    };
  \end{feynman} 
  \end{tikzpicture}  
  \label{G4:fde}
\end{equation}
Each vertex of the Feynman diagram corresponds to $-\lambda$, and each external line of the 1PI diagram contains 
wavelet factor $\tilde{\chi}(ak)$. In one-loop approximation, we have the following expressions, 
for the renormalized inverse propagator $\Gamma^{(2)}_{(A)}$ and renormalized vertex function $\Gamma^{(4)}_{(A)}$,
respectively:
\begin{align}
\frac{\Gamma^{(2)}_{(A)}(a_1,a_2,p)}{\tilde{\chi}(a_1 p) \tilde{\chi}(-a_2 p)} &= p^2+m^2 + 
\frac{\lambda}{2} T^d_\chi(A) \label{g2l1},
\\
\frac{\Gamma^{(4)}_{(A)}}{\tilde{\chi}(a_1p_1)\tilde{\chi}(a_2p_2)\tilde{\chi}(a_3p_3)\tilde{\chi}(a_4p_4) } &= \lambda -\frac{3}{2}\lambda^2 X^d_\chi(A), \label{g4l1}
\end{align}
where $A$ is the minimal scale of all external lines of the corresponding diagram.

The tadpole integral in Eq.\eqref{g2l1}, 
$$
T^d_\chi(A) = \int \frac{d^dq}{(2\pi)^d} \frac{f^2_\chi(Aq)}{q^2+m^2}$$
determines the contribution of all fluctuations with scales from $A$ to $\infty$ to the 'dressed mass' at 
the observation scale $A$. In the local theory with $\phi^4$ interaction, the natural length scale is the bare mass, the parameter of the action \eqref{se4}. Expressing  the momenta in the units of 
mass $m$, we get 
\begin{equation}
T^d_\chi(A) = \frac{S_d m^{d-2}}{(2\pi)^d} \int_0^\infty f^2_\chi(Amx) \frac{x^{d-1}dx}{x^2+1}, \label{td}
\end{equation}
where $x$ is dimensionless, and $\alpha =Am$ is dimensionless scale of observation.

For $n=1$ wavelet we get 
\begin{align}\nonumber 
T^4_{\chi_1}(A) &= \frac{m^2}{8\pi^2}\int_0^\infty e^{-2\alpha^2x^2} \frac{x^3dx}{x^2+1} \\
& = \frac{m^2}{32\pi^2} \left(\frac{1}{\alpha^2} - 2 e^{2\alpha^2} \Ei_1(2\alpha^2) \right),
\end{align}
where $\alpha = Am$ and $$\Ei_1(z):= \int_1^\infty \frac{e^{-xz}}{x}dx$$  is the exponential integral of the 
first kind.

Similarly, for the $n=2$ wavelet we get 
\begin{widetext}
\begin{align}
T^4_{\chi_2}(A) = \frac{m^2}{8\pi^2}\int_0^\infty e^{-2\alpha^2x^2} (1+\alpha^2x^2)^2 \frac{x^3dx}{x^2+1} 
= \frac{m^2}{32\pi^2} \Bigl(
\frac{5}{2\alpha^2} -\frac{5}{2} +\alpha^2 
+ 2 e^{2\alpha^2} \Ei_1(2\alpha^2) 
[2\alpha^2 -\alpha^4-1] 
\Bigr),
\end{align}
\end{widetext}
For small scales ($Am\!\ll\!1$) the one-loop contribution to the effective mass in \eqref{g2l1} is dominated by the square 
term $\propto \frac{\lambda}{A^2}$. 

Similarly, the one-loop contribution to the vertex function is given by the ''fish'' 
integral,  
\begin{equation}
X^d_\chi(A) = \int \frac{d^dq}{(2\pi)^d}
\frac{f^2_\chi(qA)f^2_\chi((q-s)A)}{\left[ q^2+m^2\right]\left[ (q-s)^2+m^2\right] }.
\label{li1}
\end{equation}
Let us consider one-loop integrals \eqref{td} and \eqref{li1} in $d\!=\!4$ dimension, where the coupling 
constant $\lambda$ is dimensionless, for different mother wavelets $n=1,2$ of the  family \eqref{gn}.

The ''fish'' integral contribution \eqref{li1} to the vertex function \eqref{g4l1} can be evaluated by symmetrization 
of loop momenta $q \to q + s/2$, where $s=p_1+p_2$ is the sum of the incoming momenta. In terms of the dimensionless 
momentum $\vy = \vq / |s|$, the integral \eqref{li1} takes the form 
\begin{widetext}
\begin{equation} 
X^d_\chi(A) = \frac{S_{d-1}s^{d-4}}{(2\pi)^d} \int_0^\pi d\theta \sin^{d-2}\theta \int_0^\infty dy y^{d-3}
\frac{f^2_\chi\left(As \sqrt{y^2+y\cos\theta + \frac{1}{4}}\right)f^2_\chi\left(As \sqrt{y^2-y\cos\theta + \frac{1}{4}}\right)
}{
\left[\frac{y^2+\frac{1}{4}+\frac{m^2}{s^2}}{y}  +\cos\theta\right]\left[\frac{y^2+\frac{1}{4}+\frac{m^2}{s^2}}{y}  -\cos\theta\right]
},
\label{Xd}
\end{equation}
\end{widetext}
where $\theta$ is the angle between the loop momentum $q$ and the total momentum $s$. 

The integral \eqref{Xd} can be evaluated in the relativistic limit $s^2\gg4m^2$. In logarithmic dimension 
$d=4$, where the coupling constant $\lambda$ is dimensionless, relativistic approximation drastically simplifies 
the integral: the dependence on the total momentum $s$ is manifested only through the dimensionless scale $As$ in wavelet 
cutoff factors $f^2_\chi$. 

For $n=1$ this gives 
\begin{align}
X^4_{\chi_1}(A) = \frac{1}{16\pi^2}\Bigl[ 
2\Ei_1(2\alpha^2)  - \Ei_1(\alpha^2)  +e^{-\alpha^2}\frac{1 - e^{-\alpha^2} }{\alpha^2}
\Bigr], \label{X41}
\end{align}
where $\alpha=As$.
Similarly, for $n=2$ we have 
\begin{widetext}
\begin{align} 
X^4_{\chi_2}(A) = \frac{1}{16\pi^2}\Bigl[ 
2\Ei_1(2\alpha^2)  - \Ei_1(\alpha^2) -  e^{-2\alpha^2} \left(\frac{5}{2\alpha^2}+\frac{1}{2} \right) 
 + e^{-\alpha^2} \bigl(
\frac{67}{128} + \frac{9}{128}\alpha^2 + \frac{1}{256}\alpha^4 + \frac{5}{2\alpha^2}\bigr)
 \Bigr] \label{X42}
\end{align}
\end{widetext} 

The details of integral evaluation can be found in the Appendix of \cite{AR2020}.

Since equation \eqref{g4l1} gives an exact (in one-loop approximation) contribution of all fluctuations 
with scales from $A$ to infinity to 
the dependence of the effective coupling constant on the observation scale $A$, this dependence 
can be written as a function of the dimensionless scale $\alpha=As$. These dependences, 
calculated with $\chi_1$ wavelet \eqref{X41} and with $\chi_2$ wavelet \eqref{X42}, respectively, are 
\begin{widetext} 
\begin{align}
\lambda_{eff}^{(1)}(\alpha^2) &= \lambda + \frac{3}{2}\frac{\lambda^2 e^{-\alpha^2}}{16\pi^2}\Bigl[
e^{\alpha^2} (2\Ei_1(2\alpha^2) 
-\Ei_1(\alpha^2)) 
+\frac{1-e^{-\alpha^2}}{\alpha^2}\Bigr], \label{rgr} \\
\nonumber 
\lambda_{eff}^{(2)}(\alpha^2) &= \lambda + \frac{3}{2}\frac{\lambda^2 e^{-\alpha^2}}{16\pi^2}\Bigl[
e^{\alpha^2} (2\Ei_1(2\alpha^2) 
-\Ei_1(\alpha^2)) 
+\frac{1-e^{-\alpha^2}}{\alpha^2} 
+ \frac{\alpha^6+18\alpha^4+134\alpha^2+384 - e^{-\alpha^2}(128\alpha^2+384)}{256\alpha^2}
\Bigr],
\end{align}
\end{widetext}
where we have changed sign in \eqref{g4l1} to invert it from $\lambda=\lambda_{bare}$ to $\lambda=\lambda_{phys}$.

\begin{figure}[ht]
\centering \includegraphics[width=8cm]{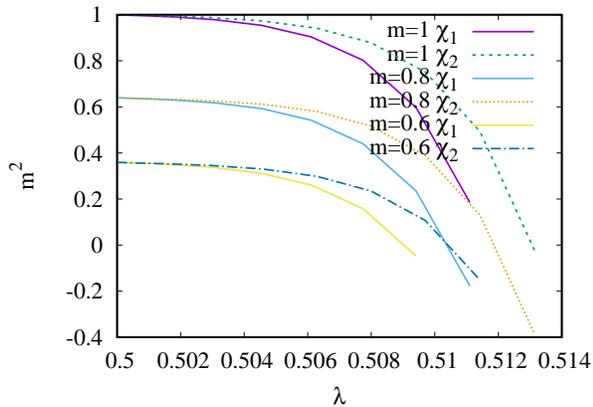}
\caption{Iteration of the finite shell renormalization (\ref{deAL},\ref{mAL}). The iteration process goes from 
$L=4$ with the value of coupling constant $\lambda=\frac{1}{2}$ at the left of the picture by setting $L_{i+1}=\sqrt{2}L_i \equiv A_i$. The right side of the picture corresponds to the UV limit of iteration. The graphs are 
shown for different values of mass and different wavelets. An arbitrary value of $s=2$ was chosen.} 
\label{rg12:pic}
\end{figure}

Let us consider the {\em contribution of a finite shell of 
scales $(A,L)$}, when a classical field is known at certain finite scale $L$, in contrast to the previous construction 
(\ref{X41},\ref{rgr}), where we have integrated out all 
fluctuations in the semi-infinite range $(A,L=\infty)$.
The value of the effective coupling constant of the type \eqref{rgr} does not diverge for any finite 
scale $A>0$ (in contrast to its differential analogue \eqref{rga1}, presented below, which suffers from the Landau pole).
The reason for this can be understood physically, if we assume a system of size $L$ in equilibrium, with well defined coupling constant $\lambda_L$. Any measurements on such system can be executed at scales $A<L$. 
The effective coupling constant relevant to a measurement at the scale $A$ is $\lambda_A$. Its particular 
value is determined by all fluctuations in the range of scales $[A,L]$. 
In one-loop approximation for the $\phi^4$ theory this effective coupling constant is 
\begin{equation}
\lambda_A = \lambda_L + \frac{3}{2}\lambda_L^2 [ X^d_\chi(A) - X^d_\chi(L) ] \label{deAL}
\end{equation}  
where the function $X(A)$ is  the ''fish'' integral of the type \eqref{X41}. 
Analogously, we can express effective mass at scale $\alpha$, using the value of physical mass $m_L$ at some large 
scale $L$ and the tadpole correction to mass,
\begin{equation}
m^2_A = m^2_L - \frac{\lambda_L}{2}(T^d_\chi(A) - T^d_\chi(L)) \label{mAL}
\end{equation}
The renormalization of the coupling constant \eqref{deAL} and the mass \eqref{mAL}, due to the integration 
over fluctuations within the range $[A,L]$, can be considered as a counterpart of the RG flow of standard 
$\phi_4^4$ model. An example, calculated from the scale $L=4$ down to the scale $A=0.0625$ with a $\sqrt{2}$ 
step in scale, is shown in Fig.~\ref{rg12:pic}.
We cannot give any comments concerning the dimension of stable and unstable manifolds of the RG flow for a number 
of reasons. First of all, in contrast to the usual RG, the case of wavelet field theory contains explicit 
dependence on the scale argument $A$. That is why we cannot draw a beta-function as a function of coupling 
constants only and find a critical point $\beta(\lambda_*)=0$ to estimate the RG flow near these points.
What we can do, instead, is to calculate the logarithmic derivatives $A \frac{\partial}{\partial A}$ of the 
coupling constants and, starting from some large scale value $\lambda_L$ of the coupling constant, iterate 
the model parameters to the smaller scales, exactly as was shown in Fig.~\ref{rg12:pic} according to 
the equations (\ref{deAL},\ref{mAL}). Needless to say, that since the physical fields in our model 
$\phi_a(x)$ explicitly depend on scale, there is no field renormalization in our model. 
Perhaps, the RG flow shown in Fig.~\ref{rg12:pic} can be considered only ''above the critical temperature'' $m^2>0$ where the wavelet evaluation of integrals is valid. Instead of considering the fixed points $\beta(\lambda_*)=0$,  
we have the explicit dependence of the coupling constant on the scale $A$. If the size of the system tends to infinity, and the physical mass $m_L$ and coupling constant $\lambda_L$ are defined in this limit, the scale 
dependence of the coupling constant $\lambda = \lambda(A)$ can be explicitly calculated from \eqref{deAL} at 
$X^d_\chi(L)=0$. The examples of this dependence are shown in Fig.~\ref{sd:pic} below.
\begin{figure}[ht]
\centering \includegraphics[width=8cm]{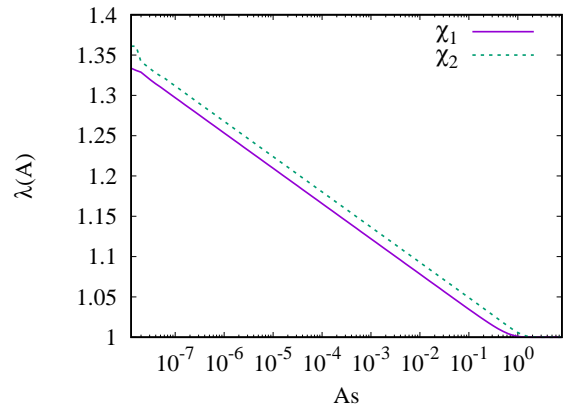}
\caption{Dependence of the coupling constant on the logarithm of the dimensionless scale $\alpha=As$ calculated 
in one-loop approximation, with $\chi_1$ and $\chi_2$ wavelets, respectively. The value of the coupling constant 
is normalized to $\lambda_L=1$ at infinity. The parameter $s=2$ was taken.}
\label{sd:pic}
\end{figure}

If the scales $A$ and $L$ 
are sufficiently close to each other, the difference equation \eqref{deAL},
$$
-\frac{\Delta \lambda}{\lambda^2}=-\frac{3}{2}\Delta X
$$
can be transformed to the differential equation $ d \frac{1}{\lambda} = -\frac{3}{2}dX$, 
which has the solution 
\begin{equation}
\lambda(A) = \frac{\lambda_L}{1 - \frac{3}{2}\lambda_L(X(A)-X(L))},
\end{equation}
which coincides with the solution of the original equation \eqref{deAL} {\em only for small values of} $\lambda_L$;
otherwise it suffers from the pole.

The formal differentiation of the effective coupling constant \eqref{rgr} with 
respect to the logarithmic scale argument gives the scaling equation 
\begin{equation}
\alpha^2 \frac{\partial \lambda}{\partial\alpha^2} = \frac{3}{2}\lambda^2\alpha^2 \frac{\partial X_{\chi_1}^4}{\partial \alpha^2}= \frac{3\lambda^2}{32\pi^2} 
\frac{e^{-\alpha^2}-1}{\alpha^2}
e^{-\alpha^2} \label{rgf}
\end{equation}
for the first wavelet $\chi_1$.
The asymptote of \eqref{rgf} for small values $\alpha\ll1$ coincides with standard result, 
$$
\frac{\partial \lambda_{eff}}{\partial\mu} \approx \frac{3\lambda^2}{16\pi^2}, \quad \mu =-\ln\alpha.
$$
Similar equations can be obtained for other wavelets of the family \eqref{gn}. For either of these wavelets,  
the asymptote of logarithmic derivatives for small scales $\alpha^2\ll1$ gives the same slope, 
$$
d_i := \alpha^2 \frac{\partial X^4_{\chi_i}}{\partial \alpha^2} = - \frac{1}{16\pi^2} + O(\alpha^2)
$$ 
for the dependence of coupling constant $\lambda$ on the logarithm of scale. For the first two 
wavelets the small scale Taylor series gives 
\begin{align*}
d_1 &= - \frac{1}{16\pi^2} + \frac{3\alpha^2}{32\pi^2} - \frac{7\alpha^4}{96\pi^2} + O(\alpha^6), \\
d_2 &= - \frac{1}{16\pi^2} - \frac{13\alpha^2}{1024\pi^2} + \frac{139 \alpha^4}{3072\pi^2} + O(\alpha^6).
\end{align*} 
The shape of the mother wavelet works as an aperture of the microscope used to study the details of 
different scales \cite{PhysRevLett.64.745}. 
Different apertures can see different values when the scale of aperture is comparable 
to size of the object, but the asymptote at small scales is certainly the same -- since the cutoff function \eqref{fcut} calculated for the \eqref{gn} family is an exponent multiplied by a polynomial $1+O(x^2)$ -- and coincides 
with the standard RG result; see also \cite{AR2020} for a similar result in quantum electrodynamics.

In the limit of small $\alpha$, the RG equation for the coupling constant, 
\begin{equation}
\frac{\partial \lambda}{\partial \ln\alpha} = - \frac{3\lambda^2}{16\pi^2}
\end{equation}
has a well-known solution, 
\begin{equation}
\lambda(\alpha) = \frac{\lambda_1}{1 + \frac{3\lambda_1}{16\pi^2} \ln \frac{\alpha}{\alpha_1}},
\label{rg0}
\end{equation}
where $\lambda_1 \equiv \lambda(\alpha_1)$ is a reference value of the coupling constant at a certain reference 
value $\alpha_1$. The solution \eqref{rg0} suffers from a Landau pole.
 
In the full form, the ordinary differential equation \eqref{rgf} can be solved as an RG-type equation, 
\begin{equation}
d \left(\frac{1}{\lambda} \right) = - \frac{3}{32\pi^2} \frac{e^{-\alpha^2}(e^{-\alpha^2}-1)}{\alpha^4} d\alpha^2
\label{rga}
\end{equation}
If the value of the effective coupling constant $\lambda$ is known at certain squared dimensionless scale 
$\lambda_1 = \lambda(x_1\!=\!(A_1s)^2)$, its value at other scales $x\!=\!(As)^2$, is given by  
the explicit solution, 
\begin{equation}
\lambda(x) = \frac{1}{\frac{1}{\lambda_1} + \frac{3}{32\pi^2} \left[F(x)-F(x_1) \right]},
 \label{rga1}
\end{equation}
where $F(x):= 2\Gamma(-1,2x)-\Gamma(-1,x)$, with $$\Gamma(a,z)=\int_z^\infty t^{a-1}e^{-t} dt$$
being the incomplete gamma-function. Similar to the small-scale case, the solution \eqref{rga1} also suffers 
from the Landau pole.

In the actual sense of the Ginzburg-Landau model, we cannot really assert that $\phi^4(x)$ interaction is 
a realistic large-scale interaction from which one can derive the small scale interaction of fields 
at $A\to0$ by means of RG and loop corrections to the large-scale theory. Instead, what we can do is to 
approximate some medium-scale interaction from the known parameters of the Hamiltonian at microscopic scale, 
i.e., at the UV-cutoff scale. In the case of a ferromagnetic model, this is the grid size. From this microscopic 
interaction, we can infer the interaction strength $\lambda$ for bigger (Kadanoff's) blocks, but not for 
the ground state of the whole crystal of finite size \cite{Wilson1971a}. In this case, the Ginzburg-Landau model is not 
a fairly good approximation.

However, there are QFT models in which the large-scale fields provide a good approximation 
for the measured physical fields, and the renormalization group with the loop corrections provides a good 
estimation of field interactions at smaller scales. Quantum electrodynamics is a well known example.

\section{Conclusions}
We have shown in this paper that the summation of all fluctuations with scales from the size 
of the system down to the observation scale by means of continuous wavelet transform results in 
a finite renormalization of the coupling constant without any Landau poles. It was demonstrated 
on a simple example of $\phi^4$ field theory in $d=4$ dimensions. Our conclusion seems rather general, since the 
same technique can be applied to QED \cite{AR2020}, QCD \cite{Altaisky2020PRD}, and other models. 
The same method of summing up the fluctuations from IR scale -- the size of the system -- down to the 
observation scale can be also applied in other dimensions, different from the dimension of physical spacetime, 
but the calculations may be more difficult technically. The one-loop integrals for the case of $\phi^4$ 
theory in $d=3$ dimensions are presented in Appendix for completeness. 

The essence of this paper is to show that that the extension of the functional space of quantum fields from 
the space of square-integrable local functions $\phi(x)$ to the space of scale-dependent functions $\phi_a(x)$, 
defined on the affine group with the help of continuous wavelet transform, leads to a theory finite by construction.
The singularities -- Landau poles, UV and IR divergences -- turn to be the artifacts of $\mathrm{L}^2(\R^d)$ space 
of functions, which is too poor to provide a correct account of the dependence of physical fields 
on the observer's settings, i.e., on the observation scale. 
Specially, the Landau poles then remain the  artifacts of approximation of the results in a {\em finite} range 
of scales by the results obtained from differential equation defined in an infinitesimally thin shell.

In probabilistic sense the summation of all fluctuations from large scale down to smaller scales 
may be related to the probabilities of small-scale fluctuations constrained by the fluctuations 
of larger scales \cite{MOBM2022}.

\section*{Acknowlengement} 
M.H. acknowledges the support from the project VEGA 1/0535/21 of Ministry of Education, Science, Research and Sport of Slovak Republic. The authors are thankful to anonymous Referee for useful comments.
\appendix
\section{Evaluation of one-loop corrections in 3d}
The one-loop corrections (\ref{g2l1},\ref{g4l1}) can be evaluated for arbitrary dimension $d$, although the calculations may be more difficult than in $d=4$, where the coupling constant is dimensionless. Here we present 
the results for $d=3$ calculated for $\chi_1$ wavelet. 

The tadpole integral \eqref{td} can be evaluated explicitly for $d=3$ with the cutoff function \eqref{fch1} $f_{\chi_1}(x)=e^{-x^2}$:
\begin{align}
T^3_{\chi_1}(A) &= \frac{m}{2\pi^2}\int_0^\infty e^{-2A^2m^2x^2} \frac{x^2dx}{x^2+1} = \\
\nonumber &= \frac{m}{2\pi^2} \left[ 
\frac{\pi}{2} e^{2A^2m^2}\bigl( \mathrm{erf}(\sqrt{2}Am)-1 \bigr)
+ \frac{\sqrt{2\pi}}{4Am}
\right].
\end{align}

The ''fish'' integral \eqref{g4l1} in $d=3$ takes the form 
\begin{widetext}
\begin{equation}
X^3_\chi(A) = \frac{1}{(2\pi)^2s} \int_0^\pi d\theta \sin\theta  \int_0^\infty dy 
\frac{
f^2_\chi \left(As \sqrt{y^2+y\cos\theta+\frac{1}{4}} \right) 
f^2_\chi \left(As \sqrt{y^2-y\cos\theta+\frac{1}{4}} \right) 
}
{
\left[\frac{y^2+\frac{1}{4} + \frac{m^2}{s^2}}{y} + \cos\theta\right]
\left[\frac{y^2+\frac{1}{4} + \frac{m^2}{s^2}}{y} - \cos\theta\right],
} \label{X43}
\end{equation}
\end{widetext}
where $s=p_1+p_2$ is the sum of the incident momenta, and $f_\chi(\cdot)$ is cutoff function 
corresponding to the chosen wavelet. 

For the simplest case of $\chi_1$ wavelet the product of the two squared  Gaussian cutoff functions in the 
numerator gives the factor $\exp\left(-\alpha^2(4y^2+1) \right)$, where $\alpha = As$. The whole integral 
\eqref{X43}, after the change of variables $u=-\cos\theta$, takes the form
\begin{equation}
X^3_{\chi_1}(A) = \frac{1}{(2\pi)^2s} \int_{-1}^1 du \int_0^\infty dy\frac{e^{-4\alpha^2(y^2+\frac{1}{4})}}{\beta^2(y)-u^2},
\end{equation}  
where 
$$
\beta(y)= \frac{y^2+\frac{1}{4}+\frac{m^2}{s^2}}{y}>1$$
in the domain of integration in the dimensionless momentum $y$. Having integrated over the angular variable $u$ 
we get 
\begin{align}\nonumber
X^3_{\chi_1}(A) &= \frac{1}{(2\pi)^2s} \int_0^\infty \frac{dy y e^{-4\alpha^2(y^2+\frac{1}{4})}}{y^2 + \frac{1}{4}+\epsilon}\times \\
&\times \left( \ln [\bigl(y+\frac{1}{2}\bigr)^2+\epsilon] -\ln [\bigl(y-\frac{1}{2}\bigr)^2+\epsilon]\right),
\label{X3log}
\end{align}
where we have introduced a presumably small parameter $\epsilon \equiv \frac{m^2}{s^2}$. In the IR limit $\epsilon\to0$ the integral \eqref{X3log} is divergent at $y=\frac{1}{2}$, otherwise its value can be calculated 
approximately, using the Laplace method. 

Changing the integration variable $y=\frac{1}{2}+t$, we get the integral 
\begin{widetext}
\begin{align}
X^3_{\chi_1}(A) &= \frac{1}{(2\pi)^2s} \int_{-\frac{1}{2}}^\infty
\frac{dt \left(\frac{1}{2}+t \right) e^{-4\alpha^2 \left(t^2+t+\frac{1}{2} \right)}}{t^2+t + \frac{1}{2}+\epsilon} 
\ln \frac{(t+1)^2+\epsilon}{t^2+\epsilon} \equiv \frac{1}{(2\pi)^2s} \int_{-\frac{1}{2}}^\infty dt \exp(S(t,\epsilon))
\end{align}
The 'action' 
\begin{equation}
S(t,\epsilon) \equiv \ln \left(t+\frac{1}{2}\right) -\ln \left(t^2+t+\frac{1}{2}+\epsilon \right) 
-4\alpha^2 \left(t^2+t+\frac{1}{2} \right) + \ln \ln \frac{(t+1)^2+\epsilon}{t^2+\epsilon}
\end{equation} 

has a sharp maximum at $t \approx 0$. The exact value of the extremal point $t_0$ is given by the equation 
$$
\left. \frac{\partial S}{\partial t }\right|_{t=t_0}=0,$$
where 
\begin{align}
\frac{\partial S}{\partial t} = \frac{1}{t+\frac{1}{2}} - \frac{2t+1}{t^2+t+\frac{1}{2}+\epsilon} 
-4\alpha^2(2t+1) 
 + \frac{\frac{2t+2}{(t+1)^2+\epsilon}-\frac{2t}{t^2+\epsilon}}{\ln  \frac{(t+1)^2+\epsilon}{t^2+\epsilon}}
\end{align}
The decomposition of the latter equation in a power series in $t$, omitting the $O(t^2)$ order terms, leads to 
\begin{equation}
t_0 \approx \frac{
-2\alpha^2+\frac{1}{\ln\frac{1}{\epsilon}}
}
{
2+4\alpha^2 + \frac{1+ \frac{1}{\epsilon} + \frac{2}{\ln\frac{1}{\epsilon}}}{\ln \frac{1}{\epsilon}}
}
\label{t0}
\end{equation}
The estimation \eqref{t0} tends to zero for $\epsilon\to0$. The second derivative of 'action' 
\begin{align}\nonumber 
\frac{\partial^2 S}{\partial t^2} &= - \frac{1}{\left(t+\frac{1}{2} \right)^2} 
-\frac{2}{t^2+t+\frac{1}{2}+\epsilon} + \left( \frac{2t+1}{t^2+t+\frac{1}{2}+\epsilon} \right)^2 -8\alpha^2
-\left(
\frac{ \frac{2t+2}{(t+1)^2+\epsilon}-\frac{2t}{t^2+\epsilon}}{\ln \frac{(t+1)^2+\epsilon}{t^2+\epsilon}}
\right)^2 \\
& + \frac{\frac{2}{(t+1)^2+\epsilon} - \frac{(2t+2)^2}{\bigl((t+1)^2+\epsilon\bigr)^2}-\frac{2}{t^2+\epsilon} + \frac{4t^2}{(t^2+\epsilon)^2} }{\ln \frac{(t+1)^2+\epsilon}{t^2+\epsilon}}
\end{align}
The value of second derivative near the extremal point is 
\begin{equation}
S_{tt}(0,\epsilon) = -4 -8\alpha^2 - \frac{2}{\frac{1}{2}+\epsilon} + \frac{1}{\left(\frac{1}{2}+\epsilon \right)^2}
+ \frac{\frac{2}{1+\epsilon} - \frac{4}{(1+\epsilon)^2}-\frac{2}{\epsilon}}{\ln \frac{1+\epsilon}{\epsilon}}
-\frac{4}{(1+\epsilon)^2 \left[\ln\frac{1+\epsilon}{\epsilon} \right]^2}
\end{equation}
Thus the ''fish'' integral for $\phi^4_3$ theory with $\chi_1$ wavelet can be estimated as 
\begin{equation}
X^3_{\chi_1}(A) \approx \frac{1}{(2\pi)^2 s} \sqrt{\frac{2\pi}{S_{tt}(0,\epsilon)}} \exp(S(0,\epsilon)), \quad
\hbox{where\ } S(0,\epsilon) = \ln \frac{\frac{1}{2}}{\frac{1}{2}+\epsilon} -2\alpha^2 + \ln \ln \frac{1+\epsilon}{\epsilon}
\end{equation}
\end{widetext}
%

\end{document}